\documentclass[aps,pre,reprint,10pt,superscriptaddress,showpacs]{revtex4-1}
\usepackage{amsfonts} 
\usepackage{amsmath}
\usepackage{amssymb}
\usepackage{graphicx}
\usepackage{subfigure}
\usepackage{color}
\newcommand*\xbar[1]{%
  \hbox{%
    \vbox{%
      \hrule height 0.5pt 
      \kern0.5ex
      \hbox{%
        \kern-0.1em
        \ensuremath{#1}%
        \kern-0.1em
      }%
    }%
  }%
} 
\begin{document}
\title{Effects of group velocity and  multiplasmon resonances on the modulation of Langmuir waves in a degenerate plasma}
\author{Amar P. Misra}
\email{apmisra@visva-bharati.ac.in; apmisra@gmail.com}
\author{Debjani Chatterjee}
\email{chatterjee.debjani10@gmail.com}
\affiliation{Department of Mathematics, Siksha Bhavana, Visva-Bharati University, Santiniketan-731 235,  India}
\author{Gert Brodin}
\email{gert.brodin@physics.umu.se}
\affiliation{Department of Physics, Ume{\aa} University, SE-901 87 Ume{\aa}, Sweden}
\pacs{52.25.Dg, 52.27.Ep, 52.35.Mw, 52.35.Sb}
\begin{abstract}
We study the nonlinear wave modulation of Langmuir waves (LWs) in a fully
degenerate plasma. Using the Wigner--Moyal equation coupled to the Poisson
equation and the multiple scale expansion technique, a modified nonlocal
nonlinear Schr{\"{o}}dinger (NLS) equation is derived which governs the
evolution of LW envelopes in degenerate plasmas. The nonlocal nonlinearity
in the NLS equation appears due to the group velocity and multi-plasmon
resonances, i.e., resonances induced by the simultaneous particle absorption
of multiple wave quanta.  We focus on the regime where the   resonant velocity of electrons is larger than the Fermi velocity and thereby the linear Landau damping is forbidden. As a result, the nonlinear wave-particle resonances due to the
group velocity and multi-plasmon processes  are the dominant
mechanisms for wave-particle interaction. It is found that in contrast to
classical or semiclassical plasmas, the group velocity resonance does not
necessarily give rise the wave damping in the strong quantum regime where $
\hbar k\sim mv_{F}$ with $\hbar$ denoting the reduced Planck's constant, $m$ the electron mass and $v_F$ the Fermi velocity, however, the three-plasmon process plays a dominant
role in the nonlinear Landau damping of wave envelopes. In this regime, the
decay rate of the wave amplitude is also found to be higher compared to that in the modest quantum
regime where the multi-plasmon effects are forbidden.
\end{abstract}
\maketitle
\section{Introduction} \label{sec-intro}
 Recently, there has been growing and considerable interest
in investigating new aspects of Landau damping in plasmas. Though, much
attention has been paid to the classical regimes (see e.g., Refs. \onlinecite
{Nicholsson,Manfredi-1997,Danielson2004}), there are many aspects which are still
unexplored in quantum regimes. When the plasma density is increased, 
various quantum effects enter the picture \cite{rightley2016, mendonca2016, daligault2014}. 
This includes e.g., the degeneracy effects \cite{rightley2016} and suppression of 
classical particle trapping due to quantum effects \cite{daligault2014}. 
Moreover, in a quantum plasma, we note that the photon momenta may be described using a distribution function,  leading to the concept of photon Landau damping \cite{mendonca2016}. 
\par
 In the well-known standard theory of Landau damping,
particles traveling  with a speed $v$ close to the phase velocity $v_p$ of a wave (i.e., $v\simeq
v_p$) feel almost a constant electric field, leading to a systematic
acceleration of particles which results into an energy transfer from waves
to resonant particles. A similar effect is also associated with the
ponderomotive force acting on charged particles in an oscillating
electromagnetic field. When these particles have a velocity close to the
group velocity $\lambda$ of wave envelopes (i.e., $v\simeq \lambda$), the accelerating
field due to the ponderomotive force changes very slowly on the same time
scale (to be more specific) as that for the evolution of wave envelopes.
This leads to a magnified energy exchange between the propagating waves and
the particles that are resonant with the ponderomotive force. We refer to
this phenomenon as the group velocity resonance. 
\par
 The group velocity resonance can be important in the diffusion of particles
in velocity space (e.g., thermalization, heating and acceleration),
transport of particle, momentum and energy. Also, since in the nonlinear
evolution of wave envelopes, the transformation of wave energy from
higher-frequency side bands to the lower-frequency side bands takes place
due to this resonance process, there may be possibility of the onset of
weak/strong turbulence in nonlinear dispersive media \cite{misra2011}. 
\par 
 Many of the more well-known aspects of Landau damping can be studied even for an
infinite plane wave. Generalizing this setup to the more realistic case of 
propagating wave packet, the effects of group velocity resonances enter the
picture \cite{ikezi1971,chatterjee2015,chatterjee2016}. In this context, the nonlinear theory of Landau damping due to group velocity
resonance of wave envelopes has  been developed in both classical 
\cite{chatterjee2015} and semiclassical plasmas \cite{chatterjee2016}.
However, it has also been shown that going beyond the linear theory, wave
damping can also take place due to multi-plasmon resonances 
\cite{brodin2016,brodin2017}, which can be present even for an infinite plane wave.
\par
In the present work, we start with the Wigner-Moyal equation, which accounts for the particle
dispersive quantum properties (but ignores other quantum effects such as
 exchange effects \cite{Ekman 2015}), coupled to the Poisson equation.
While the general theory is applicable for an arbitrary degree of
degeneracy of plasmas, we will, for simplicity, focus on the case of a fully degenerate
plasma. A generalization of our results to the  case of nonzero
electron temperature can be made with the help of results from e.g., Ref. 
\onlinecite{rightley2016}. In the fully degenerate system, the quantum
effects can be seen to appear in a wide range of wave-number scales
including, in particular, the weak quantum, the modest quantum and the
strongly quantum regimes. In the weak quantum regime in which the Langmuir
wavelength is much larger than the typical de Broglie wavelength, the
particle's resonant velocity still approaches the phase velocity of the wave
as in classical \cite{chatterjee2015} and semiclassical \cite{chatterjee2016}
theories. However, in the modest or strong quantum regimes, the resonant
velocity in the linear theory is shifted due to finite momentum and energy
of particles, i.e., $v_{res}=\omega /k\pm \hbar k/2m\equiv v_p\pm
v_{q} $, where $\omega ~(k)$ is the wave frequency (number), $v_p=\omega/k$ is the phase velocity, $h=2\pi \hbar $
is the Planck's constant,   $m$ is the electron mass and   $v_q=\hbar k/2m$ is the velocity associated with the plasmon quanta. It has been shown that (see, e.g., Ref. \onlinecite{eliasson2010}) in a fully degenerate
plasma with the background distribution as corresponding to the
zero-temperature Fermi-Dirac equilibrium in which particles can have maximum
velocity $v_{F}$, the Fermi Velocity, the linear Landau damping takes place
for particles having velocity, $v_{res}\leq v_{F}$ when $k>k_{cr}$. Here, $k_{r}$ is a
critical value to be determined from the linear dispersion relation  \cite{eliasson2010,krivitskii1991}.
\par
The scenario changes significantly when one looks for the nonlinear
evolution of waves. It has been shown in Refs. \onlinecite{brodin2016,brodin2017}
that not only one plasmon resonances take place, there are also the
possibilities of multi-plasmon resonances with velocities $v_{res}^{n}=v_p\pm nv_{q}$, where $n=1,2,3,...$, respectively, correspond to the  one-plasmon (linear), two-plasmon,
three-plasmon resonances etc. Since the results for the one-plasmon
resonance are known from the linear theory, we are, however, interested in
the resonance processes for $n>1$ in the regime of $k<k_{cr}$ in which the
linear damping (corresponding to one-plasmon resonance processes) is
forbidden.
\par
On the other hand, the nonlinear theory of electrostatic wave envelopes has been
investigated using the Vlasov-Poisson system \cite{chatterjee2015} as well
as the semi-classical limit of the Wigner-Moyal-Poisson system \cite%
{chatterjee2016}. It has been shown that the nonlinear evolution of wave
envelopes can be described by a modified nonlinear Schr{\"{o}}dinger (NLS)
equation with a nonlocal nonlinearity which appears to be due to resonant 
particles moving with the group velocity of the wave envelope. The purpose of the present
work is to consider this type of resonance as well as the multi-plasmon
resonances on the modulation and nonlinear evolution of Langmuir wave envelopes. We show
that in contrast to classical \cite{chatterjee2015} and semiclassical \cite{chatterjee2016} plasmas,  the two- and three-plasmon resonances modify the cubic nonlinearity, and moreover, the nonlocal nonlinear term is modified by the three-plasmon resonance. As a
consequence, the wave damping is significantly enhanced due to the presence of
multi-plasmon resonances that effectively convert the wave energy to the particle's 
kinetic energy.
\section{The model}\label{sec-model} 
We consider the nonlinear wave-particle interaction of
Langmuir waves in a fully degenerate quantum plasma. Our starting point is
the three-dimensional (3D) Wigner-Moyal equation for electrons, given by, 
\begin{widetext}
\begin{equation}
\frac{\partial f}{\partial t}+{\bf v}\cdot\nabla_{\bf r} f+\frac{iem^3}{(2\pi)^3\hbar^4}\int\int d^3{\bf r}' d^3{\bf v}' e^{im({\bf v}-{\bf v}')\cdot {\bf r}'/\hbar} \left[\phi\left({\bf r}+\frac{{\bf r}'}{2},t\right)-\phi\left({\bf r}-\frac{{\bf r}'}{2},t\right)\right]f({\bf r},{\bf v}', t)=0, \label{wigner-eq}
\end{equation}
\end{widetext}
where $f$ is the Wigner distribution function; $e~,m,~{\bf v}$,
respectively, are the charge, mass and velocity of electrons, and $\phi $ is
the self-consistent electrostatic potential which satisfies the 
Poisson equation 
\begin{equation}
\nabla ^{2}\phi =4\pi e\left( \int fdv-n_{0}\right) .  \label{poisson-eq}
\end{equation}%
Here, $n_{0}$ is the background number density of electrons and ions, where we for simplicity
consider an electron proton plasma. Moreover, we
consider a dense plasma with degenerate electrons in the low temperature
limit. The three-dimensional equilibrium distribution is given by 
\begin{equation}
f^{(0)}(\mathbf{v})=\left\{ 
\begin{array}{cc}
{2m^{3}}/{(2\pi \hbar )^{3}}, & |\mathbf{v}|\leq v_{F} \\ 
0, & |\mathbf{v}|>v_{F},%
\end{array}%
\right.   \label{dist-FD-3d}
\end{equation}%
where $v_{F}=\sqrt{2E_{F}/m}$ is the speed of electrons on the Fermi surface
and $E_{F}=\hbar ^{2}\left( 3\pi ^{2}n_{0}\right) ^{2/3}/2m$ is the Fermi
energy. Since we will consider the wave propagation in a single direction (which
we choose to be along the $x$-axis), it is convenient to compute the reduced
one-dimensional (1D) distribution function to be obtained by projecting the 3D distribution %
\eqref{dist-FD-3d} on the $v_{x}$-axis, i.e., using the cylindrical
coordinates in $v_{y}$ and $v_{z}$, we obtain the reduced 1D distribution
function as (replacing $v_{x}$ by $v$) 
\begin{widetext}
\begin{equation}
 F^{(0)}(v)=\int\int f^{(0)}({\bf v})dv_ydv_z=2\pi\int_0^{v_F^2-v^2}\frac{2m^3}{(2\pi\hbar)^3}u_{\perp}du_{\perp} 
 = \left\lbrace\begin{array}{cc} 
\left[{2\pi m^3}/{(2\pi\hbar)^3}\right](v_F^2-v^2),&|{ v}|\leq v_F \\
 0,&|{ v}|> v_F. 
 \end{array}\right.  \label{dist-FD-3d-reduced}
\end{equation}
\end{widetext}
\section{Derivation of the NLS equation}\label{derivation-nls}
We consider the one-dimensional propagation (along the $x$-axis) and the
 evolution of weakly nonlinear Langmuir wave envelopes in a fully
degenerate plasma. So, we introduce the multiple space-time scales as \cite
{chatterjee2015,chatterjee2016,ichikawa1974} 
\begin{equation}
\begin{split}
& x\rightarrow x+\epsilon ^{-1} \eta +\epsilon^{-2}\zeta \\
& t\rightarrow t+\epsilon ^{-1}\sigma.  \label{stretch}
\end{split}%
\end{equation}
Here, $\eta$, $\zeta$, and $\sigma$ are the coordinates stretched by a small
parameter $\epsilon$. Since the wave amplitude is infinitesimally small, so
for $t>0$ there will be a slight deviation of order $\epsilon$ from the
uniform initial value. Thus, we expand 
\begin{equation}
\begin{split}
&\phi(x,t)=\sum_{n=1}^{\infty}\epsilon^{n}
\sum_{l=-\infty}^{\infty}\phi^{(n)}_l(\eta,\sigma,\zeta) \exp[il(kx-\omega t)], \\
&f(\mathbf{v},x,t)=f^{(0)} (\mathbf{v}) \\
&+\sum_{n=1}^{\infty}\epsilon^{n} \sum_{l=-\infty}^{\infty}f^{(n)}_{l}(%
\mathbf{v},\eta,\sigma,\zeta) \exp{[il(kx-\omega t)]} ,
\end{split}
\label{expansion}
\end{equation}
where the reality conditions, namely $f^{(n)}_{-l}=f^{(n)\ast}_{l}$, $%
\phi^{(n)}_{-l}=\phi^{(n)\ast}_l$ hold. In order to properly take into
account the contributions of resonant particles, the harmonic components $%
f^{(n)}_{l}$ and $\phi^{(n)}_l$ are further expanded into Fourier-Laplace
integrals as \cite{ichikawa1974}
\begin{widetext}
\begin{equation}
\begin{split}
f^{(n)}_{l}(\mathbf{v},\eta,\sigma,\zeta)= &\frac {1}{(2\pi)^2} \int_C
d\Omega \int_ {-\infty}^{\infty}dK\tilde{f}^{(n)}_{l}(\mathbf{v}%
,K,\Omega,\zeta)  \exp[i(K\eta-\Omega\sigma)], \\
\phi^{(n)}_l(\eta,\sigma,\zeta)= &\frac {1}{(2\pi)^2} \int_C d\Omega
\int_{-\infty}^{\infty}dK\tilde{\phi}^{(n)}_l (K,\Omega,\zeta)  \exp[i(K\eta-\Omega\sigma)],  \label{Fourier-Lap-int}
\end{split}%
\end{equation}
\end{widetext}
where the contour $C$ is parallel to the real axis and lies above the
coordinate of convergence. Equation \eqref{Fourier-Lap-int} shows that the
perturbations are in the form of waves propagating along the $\eta$
-direction with a speed $\Omega/K$.
\par
We substitute the stretched coordinates \eqref{stretch}, and the expansions %
\eqref{expansion}, and \eqref{Fourier-Lap-int} into Eqs. \eqref{wigner-eq}
and \eqref{poisson-eq} to obtain, respectively,  
\begin{widetext}
\begin{eqnarray}
&&-il(\omega-kv)f^{(n)}_{l}+\frac{\partial f^{(n-1)}_{l}}{\partial \sigma}+v\frac{\partial f^{(n-1)}_{l}}{\partial \eta}+v\frac{\partial f^{(n-2)}_{l}}{\partial \zeta}\notag\\
&&+ \frac{em}{2i\pi\hbar^2}\int\int dx' d^3{\bf v}' \exp\left[im(v-v')x'/\hbar\right] \phi^{(n)}_lf_0\left\lbrace \exp\left(ik x' l/2\right)-\exp\left(-ikx' l/2\right)\right\rbrace \notag\\
&&+\frac{em}{2i\pi\hbar^2}\int\int dx' d^3{\bf v}' \exp\left[im(v-v')x'/\hbar\right] \sum_{s=1}^\infty \sum_{l'=-\infty}^\infty \phi^{(n-s)}_{l-l'}f^{(s)}_{l'}\left\lbrace \exp\left[ikx'(l-l')/2\right]-\exp\left[-ikx'(l-l')/2\right]\right\rbrace\doteq0, \label{wignermoyal1}
\end{eqnarray}
\begin{equation}
(lk)^2\phi^{(n)}_l-2ilk\frac{\partial}{\partial \eta}\phi^{(n-1)}_l -i2lk 
\frac{\partial }{\partial \zeta}\phi^{(n-2)}_l -\frac{\partial^2}{\partial
\eta^2}\phi^{(n-2)}_l-4\pi e\int f^{(n)}_{l}d^3\mathbf{v}=0,
\label{poisson1}
\end{equation}
\end{widetext}
 where the symbol $\doteq$ is used to denote the equality in
the weak sense, and we have removed the terms which contain $\phi^{(n-3)}_l$ and $\phi^{(n-4)}_l$ in Eq. \eqref{poisson1}. In the
subsequent analysis, we determine the contributions of the resonant
particles  by solving the $\sigma$-evolution
of the components $f^{(n)}_{l}$ and $\phi^{(n)}_l$ as an initial value
problem with the initial condition 
\begin{equation}
f^{(n)}_{0} (v, \eta, \sigma=0, \zeta)\doteq0,~~~ n\geq 1,  \label{int-cond1}
\end{equation}
in the multiple space-time scheme corresponding to that on the distribution
function 
\begin{equation}
f^{(n)}_{0} (v, t=0)=0.  \label{int-cond2}
\end{equation}
\subsection{Harmonic modes with $n=1,~l=1$: Linear dispersion law}
From Eqs. \eqref{wignermoyal1} and \eqref{poisson1}, equating the
coefficients of $\epsilon$ for $n=1,~l=1$, we obtain the  linear
dispersion law: \begin{widetext} 
\begin{equation}
D(k,\omega)\equiv 1+\frac{m\omega_p^2}{n_0 \hbar k^3}\int_C\frac{ 
f^{(0)}\left(\mathbf{v}+\mathbf{v}_q\right)-f^{(0)}\left(\mathbf{v} 
- \mathbf{v}_q\right)}{v_p-v}d^3\mathbf{v}=0.  \label{dispersion}
\end{equation}
\end{widetext} 
Equation \eqref{dispersion} can be rewritten as
\begin{equation}
1-\frac{\omega_p^2}{n_0k^2}\int_C \frac{f^{(0)}(\mathbf{v})}{(v_p-v)^2- 
v_q^2}d^3\mathbf{v}=0,
\end{equation}
which, in one-dimensional geometry with the reduced distribution function \eqref{dist-FD-3d-reduced}, gives 
\begin{equation}
1-\frac{\omega_p^2}{n_0k^2}\int_C \frac{F^{(0)}(v)}{(v_p-v)^2-v_q^2}dv=0.  \label{dispersion-modified}
\end{equation}
Here,  $C$ is the contour parallel to the real axis which do not need to consider the poles at $v=v_p\pm v_q$. Such an omission of the pole contribution is
due to the fact that we are interested in the regime where the one-plasmon
resonance is forbidden. Thus, to evaluate the integral in Eq. %
\eqref{dispersion-modified}, we consider only the principal value which
excludes the poles at $v=v_p\pm v_q$.
\par
Next, considering the harmonic modes for $l\neq 0,~n=1$ and the zeroth harmonic
modes for $n=1,~2;~l=0$, we obtain the following conditions: 
\begin{equation}
f^{(1)}_{\alpha,l}\doteq 0~\text{and}~\phi^{(1)}_l=0~\text{for}~|l|\geq 2,
\label{cond-f-phi-l-1}
\end{equation}
together with the zeroth-order components, given by, 
\begin{equation}
f^{(1)}_{\alpha,0}\doteq 0,~~~\phi^{(1)}_0=0.  \label{f-phi-n1-l0}
\end{equation}
\subsection{Modes with $n=2,~l=1$: Group velocity}
For $n=2,~l=1$, we have from Eqs. \eqref{wignermoyal1} and \eqref{poisson1}
the following compatibility condition for the group velocity: 
\begin{equation}
\left\{ \frac{\partial}{\partial \sigma} + \lambda \frac{\partial}{\partial
\eta}\right\} \phi^{(1)}_1(\eta, \sigma;\zeta) =0,  \label{vg-eq}
\end{equation}
where $\lambda\equiv{\partial \omega}/{\partial k}$ is the group velocity,
given by, $\lambda=\lambda_1/\lambda_2$ with 
\begin{equation}
\begin{split}
\lambda_1=2-\frac{4\pi e^2}{mk^2}\int_C \frac{v_p^2-v^2+v_q^2}{{\left\lbrace (v_p-v)^2-v_q^2
\right\rbrace}^2 } F^{(0)}(v) dv, \\
\lambda_2=-\frac{8\pi e^2}{mk^2}\int_C \frac{v_p-v}{{\left\lbrace
(v_p-v)^2-v_q^2\right\rbrace}^2 }
F^{(0)}(v) dv.  \label{lambda}
\end{split}%
\end{equation}
Equation \eqref{vg-eq} determines the $\sigma-\eta$ variation of the first
order perturbation, i.e., 
\begin{equation}
\phi^{(1)}_1 (\eta, \sigma;\zeta)=\phi^{(1)}_1 (\xi;\zeta),
\label{phi-xi-zeta}
\end{equation}
with a new coordinate $\xi$, given by, 
\begin{equation}
\xi=\eta-\lambda \sigma= \epsilon(x-\lambda t).  \label{xi-transf}
\end{equation}
It follows that the coordinate $\xi$ in Eq. \eqref{xi-transf} establishes a clear
relationship between the reductive perturbation theory and the multiple
scale expansion scheme.
\subsection{Second harmonic modes with $n=l=2$}\label{sec-2nd-harmonic}
 For the second-order quantities with $n=l=2$, we
obtain from Eqs. \eqref{wignermoyal1} and \eqref{poisson1} the 
expressions 
\begin{eqnarray}
&&f^{(2)}_2=-\frac{e}{2\hbar k(v_p-v)}\left[ \left\lbrace f_0\left(v+2v_q \right)- f_0\left(v-2v_q \right)\right\rbrace
\phi^{(2)}_2\right.  \notag \\
&&\left. +\left\lbrace f_1^{(1)}\left(v+v_q \right)-
f^{(1)}_1\left(v-v_q \right)\right\rbrace \phi^{(1)}_1\right],
\label{f2-eq}
\end{eqnarray}
\begin{equation}
\phi^{(2)}_2=-\frac{1}{8} A(k,\omega)\phi^{(1)}_1 \phi^{(1)}_1.
\label{phi-eq}
\end{equation}
The expression for $A$ is given in Appendix \ref{appendix-a}.
\subsection{Zeroth harmonic modes with $n=3,~l=0$}\label{sec-zeroth-harmonic}
 We consider the terms corresponding to $n=3,~l=0$
from Eqs. \eqref{wignermoyal1} and \eqref{poisson1}, and use Eqs. %
\eqref{f-phi-n1-l0} and \eqref{vg-eq} to obtain a set of reduced equations.
These equations are then Fourier-Laplace transformed with respect to $\eta$
and $\sigma$. Finally,  the initial condition \eqref{int-cond1} is used to obtain 
\begin{equation}
\hat{f}^{(2)}_0(v,k,\Omega,\zeta)=-\frac{K}{\Omega-Kv}\frac{e^2}{\hbar^2}%
I(v)H(K,\Omega),  \label{f20-eq}
\end{equation}
where $H(K,\Omega)$ is defined as
\begin{equation}
|\phi^{(1)}_1(\eta-\lambda \sigma,\zeta)|^2=\frac{1}{(2\pi)^2}\int d\Omega \int dKH(K,\Omega)\exp[i(K\eta-\Omega\sigma)]
\end{equation}
with
\begin{equation}
H(K,\Omega)=2\pi\delta(\Omega-K\lambda)\int dK' \phi^{(1)\ast}_1(K')\phi^{(1)}_1(K+K').
\end{equation}
The similar expression for $\hat{\phi}^{(2)}_0$ is not so required, since in the subsequent equations its coefficients appear to be vanished identically.
\subsection{Harmonic modes with $n=3,~l=1$: the NLS equation}\label{section-nls-equation}
 Finally, for $n=3$ and $l=1$ and from Eqs. %
\eqref{wignermoyal1} and \eqref{poisson1} we obtain the  modified
NLS equation 
\begin{equation}
i\frac{\partial\phi}{\partial \tau}+P \frac{\partial^2\phi}{\partial \xi^2}%
+Q |\phi|^2 \phi +\frac{R }{\pi}\mathcal{P}\int\frac{ |\phi(\xi^{\prime},\tau)|^2}{%
\xi-\xi^{\prime }} \phi d\xi^{\prime }=0,  \label{nls}
\end{equation}
for the small but finite amplitude perturbation $\phi(\xi,
\tau)\equiv\phi^{(1)}_1(\xi, \tau)$.
\par
The coefficients of the dispersion (group velocity), cubic nonlinear
(local), nonlocal nonlinear terms, respectively, are $P,~Q$ and $R$, given
by $P\equiv(1/2)\partial^2\omega/\partial k^2=\beta/ \alpha,~Q= \gamma/
\alpha$ and $R=D/ \alpha$, where 
\begin{equation}
\alpha=-\frac{8\pi e^2}{mk}\int_C \frac{v_p-v}{\left[ (v_p-v)^2-v_q^2\right] ^2} F^{(0)}(v)dv
\label{alpha}
\end{equation}
\begin{equation}
\begin{split}
\beta=&1+ \frac{4\pi e^2}{\hbar k^3}\int_C \left[\frac{\left(v-v_q
-\lambda \right)^2}{\left(v_p-v+v_q\right)^3}\right. \\
&\left.-\frac{\left(v+v_q-\lambda \right)^2}{\left(v_p-v- 
v_q\right)^3}\right]F^{(0)}(v) dv  \label{beta}
\end{split}%
\end{equation}
\begin{equation}
\gamma=\left(\frac{1}{4}\frac{A A_1}{\hbar}-\frac{1}{2\hbar^2}B+C\right)k^2,
\label{gamma}
\end{equation}
\begin{widetext} 
\begin{eqnarray}
D=-\frac{4\pi e^4}{m \hbar^2k^2} \int_\gamma\left[ \delta\left\lbrace
v-\left(v_p-3v_q \right) \right\rbrace \frac{ 
v-\lambda+ v_q}{ \left(v_p-v-v_q
\right)^3 \left( v-\lambda + 2v_q\right)}+2\delta(v-\lambda)v_q \frac{ \left\lbrace \left(
v_p-v\right)^2+v_q^2 \right\rbrace }{\left\lbrace
\left( v_p-v\right)^2-v_q^2 \right\rbrace ^3 } \right]%
F^{(0)}(v) dv.  \label{D-expression}
\end{eqnarray}
\end{widetext} 
The expressions for $A,~A_1,~B,$ and $C$ in $\gamma$ with the distribution function $F^{(0)}$ are given in 
Appendix \ref{appendix-a}. However, the reduced expressions for $\alpha,~\beta$, $\gamma$ and $D$ describing the coefficients of the NLS equation $P,~Q$ and $R$ with the Fermi-Dirac distribution \eqref{dist-FD-3d-reduced} are given in Appendix  \ref{appendix-b}.
\section{Multi-plasmon and group velocity resonances}\label{sec-multi-plasmon} 
We investigate the coefficients of the NLS
equation, especially their modifications due to the quantum effects arising
those from the Wigner-Moyal equation and the background distribution of
electrons being the Fermi-Dirac distribution, as well as, the effects due to
the wave-particle resonances. We find that the integrands in the expressions of $\alpha$ and $
\beta$ do not have any pole except at $v=v^l_{res}\equiv v_p-v_q$,
which corresponds to the linear Landau damping and lies outside the regime of
interest. The linear damping can be associated with some other factor  with a
positive sign, i.e., at  $v=v_p+v_q$. However, these are also of
less importance as the lower resonant velocity gives the wave damping more
easily. Thus, $\alpha$ and $\beta$, and hence the group velocity dispersion $%
P$ do not have any resonance contribution in the regime of interest. The
detailed discussion about the parameter regimes is given in the Subsection %
\ref{sec-parameter-regimes}.
\par
On the other hand, inspecting the local nonlinear coefficient $Q$ of the NLS
equation \eqref{nls}, and looking at the denominators of different
expressions for $A,~A_1,~B$ and $C$  (see Appendix \ref{appendix-a})  in $Q$ and factorizing them we find that 
\begin{equation}
\left(\omega-kv\right)^2-k^2v_q^2=\left(\omega-kv-kv_q\right)\left(\omega-kv+kv_q\right),
\end{equation}
\begin{equation}
\left(\omega-kv-kv_q\right)^2-k^2v_q^2
=\left(\omega-kv-2kv_q\right)(\omega-kv),
\end{equation}
\begin{equation}
\begin{split}
\left(\omega-kv-kv_q\right)^2-4k^2v_q^2
=&\left(\omega-kv-3kv_q\right) \\
&\times\left(\omega-kv+kv_q\right).
\end{split}%
\end{equation}
Thus, in addition to the resonance at the phase velocity $(v=v_p)$ and
the linear resonance $(v=v_{res}^l$), the two- and three-plasmon
resonances also occur for $\omega-kv\pm nkv_q=0$, i.e., at $v_{res}^n=v_p-nv_q$ for $n=2,3$. The other resonant velocities for $n=4,5,...$ will not appear as those are associated with higher orders of $%
\epsilon$ than cubic, which is not the present case. Note that the
resonances at $v=v_p$, $v=v^l_{res}$ and at $v^n_{res}=v_p+ nv_q $ are not of interest in the present study as those fall in the regime
of $k>k_{cr}$, where $k_{cr}$ is some critical value of $k$ which can be
shown to be $\lesssim1$ in the strong quantum regime where $%
\hbar\omega_p\sim mv_F^2$ \cite{eliasson2010,krivitskii1991}. In fact, we have the relation
for the resonant velocities $v_p-3v_q<v_p-2v_q<v_p-v_q<v_p<v_p+nv_q$. Thus, the nonlinear coefficient of the NLS equation
is significantly modified by the resonance contributions from the two- and
three-plasmon processes, which do not appear in classical \cite%
{chatterjee2015} or semiclassical \cite{chatterjee2016} plasmas. In what
follows, looking at the nonlocal coefficient $R\propto D$, we find in $D$
that the first term is the contribution from the three-plasmon resonance,
while the second one is from the group velocity resonance.
\par
Thus, from the above discussion, one concludes that in contrast to classical 
\cite{chatterjee2015} or semiclassical \cite{chatterjee2016} plasmas, while
the local nonlinear coefficient $Q$ contains resonance contributions from
two- and three-plasmon processes, the nonlocal nonlinear coefficient $R$ of
the NLS equation appears to be modified by the group velocity, as well as the
three-plasmon resonances. 
\section{Langmuir envelopes with zero-temperature Fermi-Dirac distribution}
We consider the amplitude modulation and the nonlinear evolution of Langmuir
envelopes in a fully degenerate plasma. The background distribution of
electrons are assumed to be given by the Fermi-Dirac distribution at zero
temperature [Eq. \eqref{dist-FD-3d-reduced}]. In this situation, the linear
dispersion law, the group velocity, as well as, the coefficients of the NLS
equation \eqref{nls} will be reduced. The dispersion relation %
\eqref{dispersion} reduces to 
\begin{widetext} 
\begin{equation}
1+\frac{3 \omega_p^2}{4 k^2 v_F^2} \left(2-\sum_{j=\pm1}\frac{j}{2v_q v_F%
}\left\lbrace v_F^2 - \left(v_p+jv_q \right)^2 \right\rbrace \log \left\vert%
\frac{v_p+jv_q-v_F}{v_p+jv_q+v_F}\right\vert\right) =0,
\label{dispersion-reduced}
\end{equation}
\end{widetext} where $\omega_p=\sqrt{4\pi e^2n_0/m}$ is the electron plasma
oscillation frequency.  Before we proceed to analyze the modulational instability and
the nonlinear evolution of Langmuir waves, we first investigate some
parameter regimes of interest which may correspond to semiclassical and
quantum plasmas. These are discussed in the following subsection \ref%
{sec-parameter-regimes}.

\subsection{Parameter regimes}\label{sec-parameter-regimes}
 We investigate different parameter regimes in
the plane of particle's velocity $(v)$ and the wave number $(k)$ where the
group velocity and/or multi-plasmon resonance effects become significant. Since
we consider the region of small wave numbers, i.e., $k\lambda _{F}\lesssim 1$%
, where $\lambda _{F}=v_{F}/\omega _{p}$ is the Fermi wavelength, for which
the linear resonant velocity lies outside the background distribution, i.e., 
$v_{p}\pm v_{q}>v_{F}$, the dispersion relation \eqref{dispersion-reduced}
can further be reduced. So, expanding the $\log $ functions for small wave
numbers and keeping terms upto $o(k^{4})$ we obtain from Eq. \eqref{dispersion-reduced}  \cite{eliasson2010,ferrel1957} 
\begin{equation}
\omega ^{2}=\omega _{pe}^{2}+\frac{3}{5}k^{2}V_{Fe}^{2}+(1+\alpha )k^2v_q^2,  \label{disp-final}
\end{equation}%
where $H=\hbar \omega _{p}/mv_{F}^{2}$ is the dimensionless quantum
parameter and $\alpha =(48/175)m_{e}^{2}V_{Fe}^{4}/\hbar ^{2}\omega
_{pe}^{2}=(48/175)H^{2}$. We note that $H$ may be of the order of unity for
metallic densities, which means that typically $\alpha <1$. 
\begin{figure*}[tbp]
\includegraphics[scale=0.5]{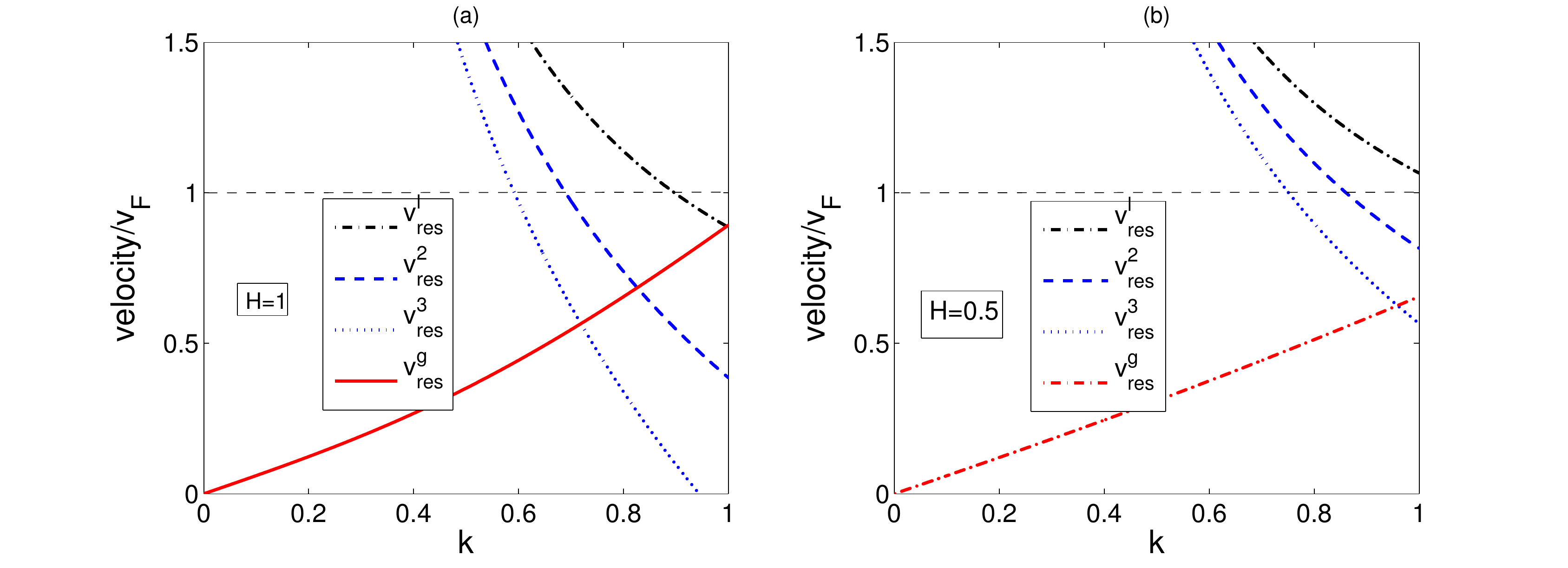}
\caption{The normalized resonant velocities ($\sim v_F$) are plotted against the normalized wave number $k~(\sim \lambda_F^{-1})$ for two different values of the dimensionless quantum parameter $H$ to show different parameter regimes, namely semi-classical (e.g., $0<k\ll0.59$ for $H=1$), modest quantum ($0<k\lesssim0.59$ for $H=1$ and $0<k\lesssim0.75$ for $H=0.5$) and strong quantum ($0.591\lesssim k\lesssim0.9$ for $H=1$ and $0.75\lesssim k\lesssim1$ for $H=0.5$) regimes. In the legends, $v_{res}^n$ denotes the resonant velocity, where $n=l,2,3,g$, respectively, correspond to the velocities for the linear, two-plasmon, three-plasmon and group velocity resonances.  }
\label{fig1}
\end{figure*}
\par
We plot the resonant velocities $v^n_{res}=v_p-nv_q,~n=l, 2,3$,
in the $vk$ plane for two different values of $H$: $H\sim1$ and $H\sim0.5$.
Here, the velocity is normalized by the Fermi velocity $v_F$ and $k$ by the
inverse of the Fermi wavelength $\lambda_F^{-1}$, and the expression for $%
v_p$ is used from Eq. \eqref{disp-final}. From Fig. \ref{fig1}, it is
clear that there are, in fact, two parameter regimes: one where both the
multi-plasmon and the group velocity resonances can be important, and the
other where the group velocity resonance is only the damping mechanism. We
note that the group velocity resonance (which occurs at $v_{res}^g=\lambda$) curve is always within
the region of $v<v_F$ throughout the interval $0<k\lesssim1$. However, the
multi-plasmon resonance curves may or may not fall in the region of $v<v_F$
depending on the values of $k$ and $H$. For $H\sim1$, subplot \ref{fig1}(a)
shows that the regime of $k$ for which the linear resonance is forbidden is $%
0<k\lesssim0.9$. However, the two- and three-plasmon resonances disappear
and only the group velocity resonance comes to the picture in $%
0<k\lesssim0.59$. On the other hand, both the three-plasmon and the group
velocity resonances can be effective in the regime $0.59\lesssim
k\lesssim0.6953$. Furthermore, the group velocity, as well as the two- and
three-plasmon resonances can be significant in $0.59\lesssim k\lesssim0.9$.
In these regimes of $k$, the magnitudes of the coefficients $P,~Q$ and $R$
of the NLS equation \eqref{nls} are to be noticed. As will be shown later, these magnitudes essentially give some useful information for the estimation of frequency shift and the rate of energy transfer in the modulation of Langmuir waves, as well as the  nonlinear evolution of envelope solitons.   For example, at $H=1$,
while the values of $P$ and $R$ increase, the values of $|Q|$ decrease with successive reduction of the values of  $k$   from $k=0.59$ to that in the regime $0<k\lesssim0.59$. In the other regime, i.e.,  $0.59\lesssim k\lesssim0.9$ with $H=1$, while the values of $P$ and $|Q|$ decrease, the values of $R$ increase
with increasing values of $k$ until the inequality $v_{res}^3\gtrsim v^g_{res}$
holds. An opposite trend occurs with $v_{res}^3<v^g_{res}$ where the
values of $P$ increase, but those of $|Q|$ and $R$ decrease with increasing
values of $k$.
\par
On the other hand, as the value of the quantum parameter $H$ is reduced
[see, e.g., subplot \ref{fig1}(b) for $H=0.5$], the multi-plasmon resonance
curves tend to disappear from the region: $0<k<1$, $v<v_F$, and they
completely disappear (not shown in the figure), e.g., at $H\sim0.1$. The latter, in some sense, corresponds to a weak quantum regime.  In this
situation, only the group velocity resonance becomes significant as in
classical or semiclassical plasmas \cite{chatterjee2015,chatterjee2016}. Thus, from the consequences of Fig. \ref%
{fig1}, one can, in particular, define three different regimes of interest:
(i) Semi-classical or weak quantum regime (ii) Modest quantum regime and
(iii) Strong quantum regime, which we briefly discuss as follows:
\begin{itemize}
\item \textit{Semi-classical regime}: If the Langmuir wavelength is much
larger than the typical de Broglie wavelength, i.e., $\hbar k\ll mv_F$, the
quantum effects associated with the terms $\propto\hbar k/m$, which appear
due to the use of the Wigner equation rather than the Vlasov equation, are
almost negligible. In this case, the quantum contributions are only due to
the background distribution of electrons being a Fermi-Dirac distribution at
zero temperature rather than a Maxwellian one. Though, the coefficients of
the NLS equation will be somewhat modified, however, the results will be similar to
some previous works \cite{chatterjee2016}, because the resonance velocity is only
the group velocity. The regimes of $k$ can be sort of $0<k\ll0.59$ for $%
H\sim1$. Furthermore, in this regime, the nonlocal nonlinear coefficient $R$
of the NLS equation, which basically modifies the shape of a pulse profile,
remains positive implying that only the group velocity resonance gives rise the
Nonlinear Landau damping of wave envelopes in the semiclassical regime.
\item \textit{Modest quantum regime}: In this case, $\hbar k\sim mv_F$,
however, the three-plasmon resonance velocity is slightly larger than the
Fermi velocity, i.e., $v_{res}^3>v_F$. This means that the resonance
contribution is still due to the group velocity [see Fig. \ref{fig1}(a)].
Though, the results may be similar to the semiclassical case, however, the
coefficients of the NLS equation will be modified by the quantum
contributions from the Wigner equation, as well as from the background
distribution of electrons. From Fig. \ref{fig1}, it is evident that the
corresponding values of $k$ are in $0<k\lesssim 0.59$ for $H\sim1$ and in $0<k\lesssim0.75$
for $H\sim0.5$. In this case, $R$ is also positive, and similar conclusion
can be drawn as for the semiclassical regime.
\item \textit{Strong quantum regime}: The most important and interesting is the strong
quantum regime where $\hbar k\sim mv_F$ still holds, however, the three-plasmon
resonance velocity is smaller than the Fermi velocity, i.e., $v_{res}^3<v_F$%
. In this case, not only the group velocity resonance contributes, but also
the two- and three-plasmon resonances come to the picture as is evident from
Fig. \ref{fig1} (see dashed and dotted lines). Also, we note that the three-plasmon resonance contribution
to the nonlocal coefficient $R$ is proportional to the difference between $%
v_F$ and $v_{res}^3$. Furthermore, the resonance contributions from the two-
and three-plasmon processes in the local nonlinear coefficient $Q$ are also
proportional to the difference $v_F-v^n_{res}$ (for $n=2,3$). Thus, the effects of the  two-
and three-plasmon resonances are to be important, the inequality $v_F>v^n_{res}$
must hold at least with a small margin. In this case, $R$ remains not only
positive in $0.591\lesssim k\lesssim0.9$, but also the contribution from the
three-plasmon resonance becomes higher in  magnitude as long as $v_{res}^3~(<v_F)$ remains
close to but slightly larger than $v^g_{res}$. Thus, it follows that in the
strong quantum regime, the group velocity resonance does not necessarily
play a decisive to the wave damping as in the semiclassical and modest
quantum regimes, however, the three-plasmon resonance plays a dominating
role in the Landau damping process.
\end{itemize}

\section{The nonlinear landau damping and modulational instability}\label{sec-MI} 
We consider the amplitude modulation of Langmuir wave envelopes in a
degenerate plasma. To this end, we assume a plane wave solution of Eq. %
\eqref{nls} of the form 
\begin{equation}
\phi= \rho^{1/2}\exp\left(i \int ^\xi \frac{\sigma}{2P} d\xi\right),
\label{sol-nls}
\end{equation}
where $\rho$ and $\sigma$ are real functions of $\xi$ and $\tau$.
Substitution of the solution \eqref{sol-nls} into Eq. \eqref{nls} results
into a set of equations which can be separated for the real and imaginary
parts. These equations are then linearized by splitting up $\rho$ and $%
\sigma $ into their equilibrium (with suffix $0$) and perturbation (with
suffix $1$) parts, i.e., 
\begin{equation}
\rho= \rho_0 +\rho_1 \cos{(K\xi-\Omega \tau)}+\rho_2\sin{(K\xi-\Omega \tau)},
\label{rho-perturbation}
\end{equation}
\begin{equation}
\sigma= \sigma_1 \cos{(K\xi-\Omega \tau)}+\sigma_2 \sin{(K\xi-\Omega \tau)},
\label{sigma-perturbation}
\end{equation}
where $\Omega$ and $K$ are, respectively, the wave frequency and the wave
number of modulation, to obtain the dispersion relation \cite{ichikawa1974,chatterjee2015} 
\begin{equation}
(\Omega^2 +2\rho_0PQK^2-P^2K^4)^2=-(2\rho_0PR K^2)^2.  \label{dispersion-nls}
\end{equation}
The equation \eqref{dispersion-nls} is, in general, complex, irrespective of
the negative sign and/or the presence of the nonlocal coefficient $R$ on the
right-hand side, as $Q$ contains pole contributions from the multi-plasmon
processes. In the semiclassical and modest quantum regimes, where the
multi-plasmon resonances are forbidden for which $Q$ is real, the dispersion
relation \eqref{dispersion-nls} can be complex due to the presence of $R$ on
the right-hand side, irrespective of whether $PQ>0$ or $PQ<0$ as in the case
of an ordinary NLS equation.
\par
A general solution of Eq. \eqref{dispersion-nls} can be obtained by
considering $\Omega= \Omega_r + i\Gamma $, $Q=Q_1+iQ_2$ with $%
\Omega_r,~\Gamma,~Q_1,~Q_2$ being real and $Q_2$ is the resonance
contribution from the multi-plasmon processes, as 
\begin{equation}
\begin{split}
\Omega_r=&\pm\frac{|K|}{\sqrt{2}}\left[\left\lbrace \left(P^2
K^2-2\rho_0PQ_1\right)^2+\left[2\rho_0P\left(R+Q_2\right)\right]%
^2\right\rbrace^{1/2}\right. \\
&\left. +\left(P^2 K^2-2\rho_0PQ_1\right)\right]^{1/2},  \label{Omega-real}
\end{split}%
\end{equation}
\begin{equation}
\begin{split}
\Gamma=&\mp\frac{|K|}{\sqrt{2}}\left[\left\lbrace\left(P^2
K^2-2\rho_0PQ_1\right)^2+\left[2\rho_0P\left(R+Q_2\right)\right]%
^2\right\rbrace^{1/2}\right. \\
&\left.-\left(P^2 K^2-2\rho_0PQ_1\right)\right]^{1/2},  \label{Omega-img}
\end{split}%
\end{equation}
where we consider the upper (lower) sign for $K>0~(K<0)$.
From Eqs. \eqref{Omega-real} and \eqref{Omega-img} it is clear that, in
comparison with classical \cite{chatterjee2015} or semiclassical \cite%
{chatterjee2016} results, both the frequency shift and the  energy transfer
(from the wave energy to the particle's kinetic energy) rate get modified by the imaginary part of $Q$ associated with the multi-plasmon resonances.
\par
\begin{figure*}[ht]
\centering
\includegraphics[height=2.5in,width=6.5in]{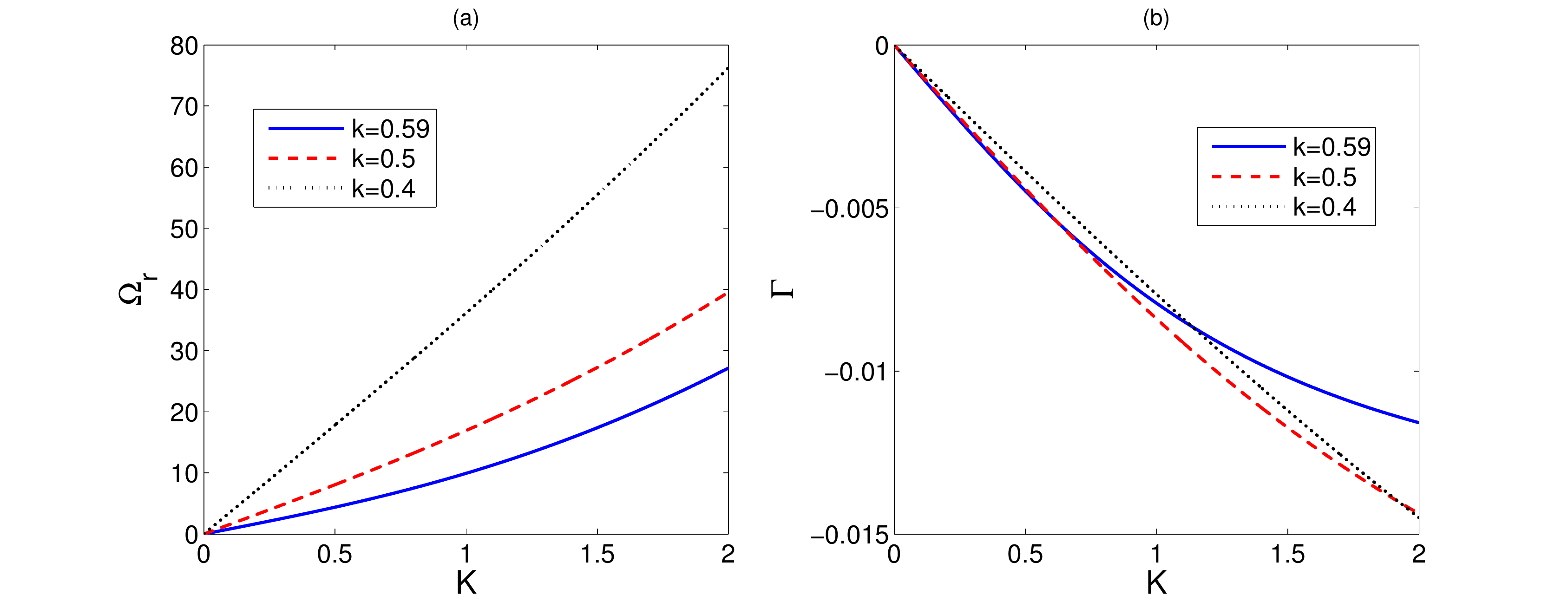}
\caption{The normalized frequency shift $\Omega_r~(\sim\omega_p)$ [Eq. 
\eqref{Omega-real}, panel (a)] and the energy transfer rate $\Gamma~(\sim
\omega_p)$ [Eq. \eqref{Omega-img}, panel (b)] are plotted against the
normalized wave number of modulation $K~(\sim \lambda_F^{-1})$ for different values of the
carrier wave number $k~(\sim \lambda_F^{-1})$ that correspond to semi-classical and modest quantum regimes. }
\label{fig2}
\end{figure*}
\begin{figure*}[ht]
\centering
\includegraphics[height=2.5in,width=6.5in]{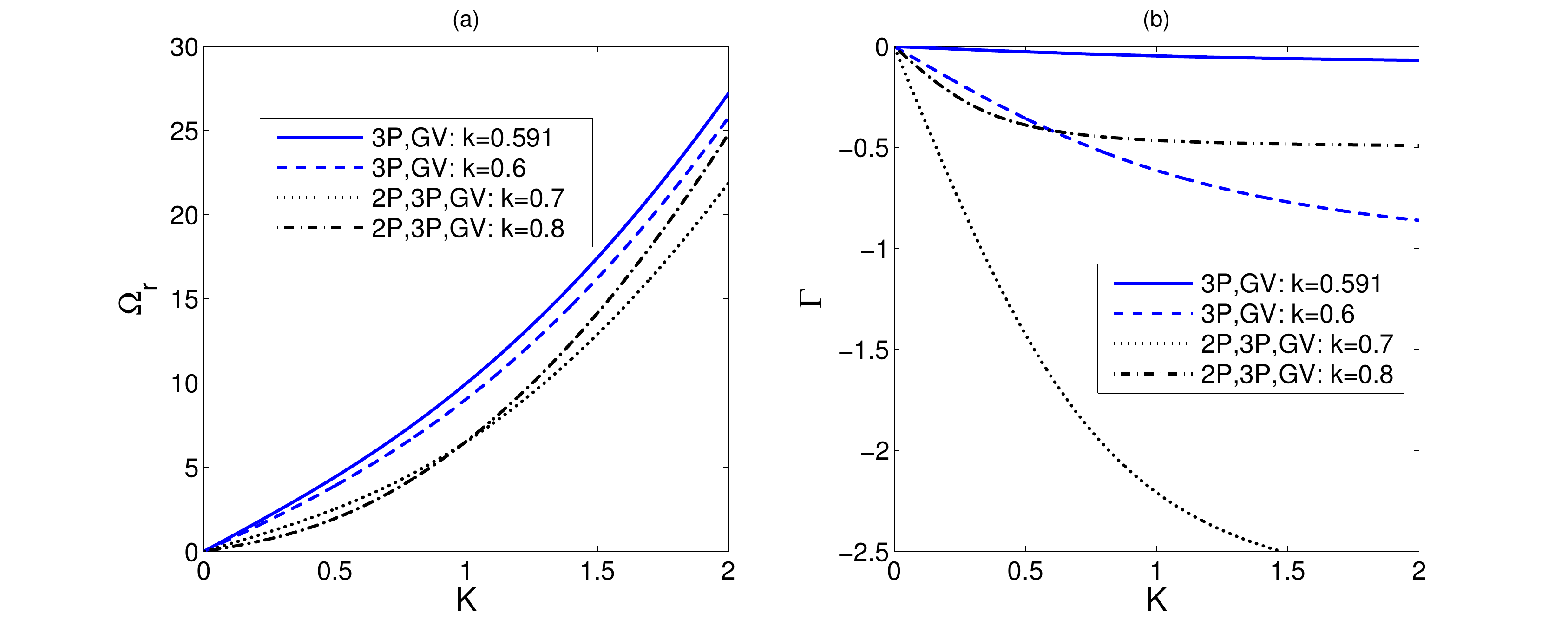}
\caption{ The same as in Fig. \protect\ref{fig3} but in the strong quantum
regime. In the legends, $2P,~3P$ and $GV$, respectively, stand for
two-plasmon, three-plasmon and group velocity resonance effects. }
\label{fig3}
\end{figure*}

In what follows, we numerically analyze the properties of $\Omega_r$ and $%
\Gamma$ for different values of the carrier wave number $k$ which
correspond to, especially   the modest and strong quantum regimes as discussed
before  (Since the semi-classical results are similar to the previous works \cite{chatterjee2016}, we skip those discussion in the present work).  The results are displayed in Figs. \ref{fig2} and \ref{fig3}. The
panel (a) in each figure shows the plots of the frequency shift $\Omega_r$,
and the other panel (b) that for the energy transfer rate $\Gamma$ against the
dimensionless wave number of modulation $K ~(\lambda_F^{-1})$. Figure \ref{fig2} shows the curves for $%
\Omega_r$ and $\Gamma$ in the semi-classical and modest quantum regimes where the imaginary
part of $Q$ is zero. We choose a value of $k=0.59$ (see the solid lines)
at which the three-plasmon resonant velocity $v_{res}^3$ marginally exceeds
the Fermi velocity $v_F$, and the effects of the two and three-plasmon
resonances are thereby forbidden. So, only the resonance effect comes from the group
velocity. The values of $k$ are then lowered from $k=0.59$ to the values $k=0.5$ and $k=0.4$ (see the dashed and dotted lines). It is found that as one approaches
the regimes of low wave numbers, the frequency shift increases gradually,
however, the values of $|\Gamma|$ increase until $k$ assumes the value $%
k=0.5 $, and then decrease as the value of $k$ is further lowered from $k=0.4$. The
reason is that at $k=0.59$, the  contribution from the cubic nonlinearity $Q$ becomes higher in magnitude than those of  the group velocity dispersion $P$ and the nonlocal nonlinearity $R$. However, as $k$ is further lowered from $k=0.59$ to $k=0.5$, the magnitude of $Q$ gets highly reduced being lower than (but comparable to) that of $P$ but still larger than $R$. As a result, both $\Omega_r$ and $|\Gamma|$ remain higher at $k=0.59$ than those at $k=0.5$.  The magnitude of  $Q$ becomes significantly reduced at $k\lesssim0.4$ in which the group velocity dispersion $P$ dominates over $Q$ and $R$.  So, in the regime of low wave numbers (below $k=0.5$), though the frequency shift remains high,   the magnitude of $\Gamma$ gets highly reduced. This implies that the rate of transfer of wave energy
  to the particle's kinetic energy may not be faster as one approaches to the semi-classical regimes where $\hbar k/mv_F\ll1$ at some small wave number $k$ is satisfied, and this transfer rate can be maximum
near the point $k=0.5$ where the three-plasmon resonant velocity slightly exceeds the Fermi velocity.
\par
The scenario changes significantly when the three-plasmon resonance effect
starts playing a role in the strong quantum regime $0.591\lesssim
k\lesssim0.9$. Figure \ref{fig3} shows that at $k=0.591$, the three-plasmon velocity is
close to but smaller than $v_F$ for which the contribution from the
three-plasmon resonance in $R$ becomes smaller than that from the group
velocity resonance. As a result the frequency shift remains high, however, $|\Gamma|$ attains its minimum value. As are seen from the solid and dashed
lines that the effect of the three-plasmon process is to decrease  the
values of $\Omega_r$ but to increase the values of  $|\Gamma|$. The similar results are also found with
the effects of the two-plasmon resonance while combined with the effects of
the three-plasmon process in the regime $0.6953\lesssim k\lesssim 0.9$ as
long as the inequality $v^3_{res}\gtrsim v^g_{res}$ holds [see the dotted lines]. In this
restriction, the values of both $P$  and $|Q|$ decrease, but those of $R$
increase. However, as we further increase the value of $k$, i.e., $k=0.8$ such that
  $v^3_{res}<v^g_{res}$ holds, the reduction of $|Q|$ becomes significantly high,   the values of   $P$ start  increasing and those of $R$ decreasing. As a result,  while the frequency shift starts increasing, the values of $|\Gamma|$ decrease with decreasing values of $k$ [see the dash-dotted lines]. Thus, it follows that the effect of the three-plasmon resonance is to reduce the frequency shift but to  enhance the energy transfer rate whenever the corresponding resonant velocity remains greater than the group velocity resonance.   
\section{Nonlinear Landau damping of solitary wave solution}\label{sec-soliton-sol} 
It is to be noted that in absence of the nonlocal coefficient $R$, the NLS
equation \eqref{nls} possesses an infinite number of conservation laws. The
first three conserving quantities are namely, the mass $I_1=\int|\phi|^2d\xi$%
, the momentum $I_2=(2i)^{-1}\int\left(\phi^{\ast}\partial_{\xi}\phi-\phi%
\partial_{\xi}\phi^{\ast}\right)d\xi$ and the wave energy $%
I_3=\int\left(|\partial_{\xi}\phi|^2-(|Q/2P|)|\phi|^4\right)d\xi$. However,
the similar quantities for the modified NLS equation \eqref{nls} with
nonzero $R$ satisfy the following equations: 
\begin{equation}
\frac{\partial I_1}{\partial\tau}=0,  \label{mass-conserv}
\end{equation}
\begin{equation}
\frac{\partial I_2}{\partial\tau}+\frac{R}{\pi}\mathcal{P}\int\int \frac{1}{%
\xi-\xi^{\prime }}|\phi(\xi^{\prime},\tau)|^2\frac{\partial}{\partial\xi}%
|\phi(\xi,\tau)|^2 d\xi d\xi^{\prime }=0,  \label{momentum-conserv}
\end{equation}
\begin{equation}
\begin{split}
\frac{\partial I_3}{\partial\tau}+&i\frac{R}{\pi}\mathcal{P}\int\int \frac{1%
}{\xi-\xi^{\prime }}|\phi(\xi^{\prime},\tau)|^2 \\
&\times\frac{\partial}{\partial\xi}\left(\phi\frac{\partial^2}{\partial\xi^2}%
\phi^{\ast}-\phi^{\ast}\frac{\partial^2}{\partial\xi^2}\phi\right) d\xi
d\xi^{\prime }=0.  \label{energy-conserv}
\end{split}%
\end{equation}
From Eq. \eqref{energy-conserv}, upon using the fact that the integral over $%
\xi$ is a convolution of the functions $\mathcal{P}\left[1/(\xi^{\prime
}-\xi)\right]$ and $\partial_{\xi}\varphi(\xi,\tau)$, where $%
\phi\partial_{\xi}^2\phi^{\ast}-\phi^{\ast}\partial_{\xi}^2\phi=
\partial_{\xi}\left(\phi\partial_{\xi}\phi^{\ast}-\phi^{\ast}\partial_{\xi}%
\phi\right)\equiv \partial_{\xi}\varphi(\xi,\tau)$, and noting that the
Fourier inverse transform of $i~\text{sgn}{(s)}=-(1/\pi)\mathcal{P}(1/\xi)$,
we obtain 
\begin{equation}
\int \frac{\partial \varphi(\xi,\tau)}{\partial\xi}\mathcal{P}\frac{1}{%
\xi^{\prime }-\xi}d\xi=\frac{1}{2}\int\exp(is\xi^{\prime })|s|\hat{\varphi}%
(s,\tau)ds.
\end{equation}
So, performing the integral over $\xi^{\prime}$ as a Fourier transform of $%
|\phi(\xi^{\prime},\tau)|^2$ we obtain 
\begin{equation}
\begin{split}
\mathcal{P}\int\int & \frac{1}{\xi-\xi^{\prime }}|\phi(\xi^{\prime},\tau)|^2\frac{
\partial \varphi(\xi,\tau)}{\partial\xi}d\xi d\xi^{\prime } \\
&=\frac{1}{2}\int|s|\hat{\varphi}(s,\tau)|\hat{\phi}(-s,\tau)|^2ds,
\end{split}%
\end{equation}
where `hat' denotes the Fourier transform with respect to $\xi$ or $%
\xi^{\prime }$. Furthermore, using $\hat{\varphi}(s,\tau)\equiv-2is|\hat{\phi%
}(s,\tau)|^2$ we obtain from Eq. \eqref{energy-conserv} 
\begin{equation}
\frac{\partial I_3}{\partial\tau}=-\frac{R}{\pi}\int s^2|\hat{\phi}%
(s,\tau)|^2|\hat{\phi}(-s,\tau)|^2ds.  \label{energy-conserv-reduced}
\end{equation}
The left-hand side of Eq. \eqref{energy-conserv-reduced} represents the rate
of change of the wave energy, and the integral on the right-hand side is a
positive definite. Thus, it follows that the wave amplitude decreases or
increases depending on whether $R>0$ or $<0$. In the former case, we have
the inequality (the equality holds for $\phi=0~\forall~\xi$) 
\begin{equation}
\frac{\partial I_3}{\partial\tau}\leq0,  \label{dI2-dtau}
\end{equation}
implying that an initial perturbation (e.g., in the form a plane wave) will
decay to zero with time $\tau$, and hence a steady state solution with $%
|I_3|<\infty$ of the NLS equation \eqref{nls} may not exist in presence of
the nonlocal term $\propto R$. In this situation, an approximate soliton
solution of the NLS equation \eqref{nls} with a small effect of the
nonlinear Landau damping $(\propto R)$ can be obtained whose  amplitude is of the
form \cite{chatterjee2015,chatterjee2016} 
\begin{equation}
\phi(\xi,\tau)\propto\sqrt{\phi_{0}(\xi,0)}\left(1-i\frac{\tau}{\tau_0}%
\right)^{-1/2},  \label{sol1-approx-nls}
\end{equation}
where $\tau_0$ is some constant inversely proportional to $R$ and $%
\phi_{0}(\xi,0)$ is the value of $\phi$ at $\tau=0$ (For details, see, e.g., Refs. \onlinecite{chatterjee2015}).
\par
A careful examination reveals that the coefficient $R$ of the NLS equation %
\eqref{nls} is always positive in the parameter regimes as shown in Fig. \ref%
{fig1}. However, the contribution from the three-plasmon resonance becomes
higher as the value of $v^3_{res}$ is gradually lowered from $v_F$ until the
relation $v^3_{res}\gtrsim v^g_{res}$ holds. A qualitative plot of the decay
rate $DR\equiv|\left(1-i\tau/\tau_0\right)^{-1/2}|$ is shown in Fig. \ref%
{fig4} in the modest and strong quantum regimes to show the relative
importance of the group velocity (solid and dashed lines) and three-plasmon
(dotted and dash-dotted lines and as indicated in the figure) resonances. We
find that in the regime where the resonance contribution is only from the
group velocity, as the wave number decreases and hence the group velocity,
the decay rate of the wave amplitude becomes higher or the magnitude of the wave amplitude gets highly reduced.
However, the decay rate can be higher or the magnitude of the wave amplitude
can be minimized in presence of the three-plasmon effect. The black-dashed
line shows that as the restriction $v^3_{res}\gtrsim v^g_{res}$ is relaxed
at a higher value of $k$, the decay rate is further reduced or the magnitude of $DR$ is increased (compare red dash-dotted and black dashed lines). This is a consequence to the fact that as the carrier wave number $k$ is increased to $k=0.8$ or above, the effects  of both  the cubic nonlinearity and nonlocal nonlinearity get significantly diminished, however,  those of the group velocity dispersion are enhanced. 
\begin{figure*}[ht]
\centering
\includegraphics[height=2.5in,width=6.5in]{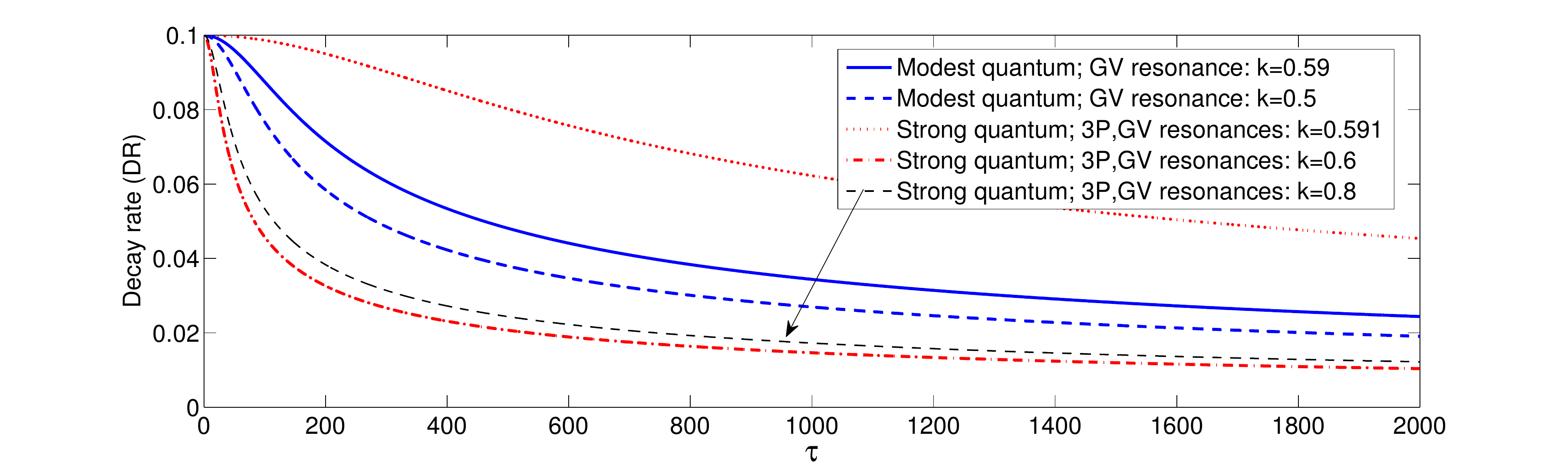}
\caption{The absolute value of the decay rate $DR\equiv|\left(1-i\protect\tau%
/\protect\tau_0\right)^{-1/2}|$ is shown against the normalized time
variable $\tau~(\omega_p^{-1})$ in different parameter regimes as in
the legend. }
\label{fig4}
\end{figure*}
\section{Discussion and Conclusion}
We have investigated the nonlinear wave modulation of Langmuir waves in a
fully degenerate plasma. Starting from the Wigner-Moyal equation coupled to
the Poisson equation and using the multiple scale expansion technique we
have derived a modified NLS equation with a nonlocal nonlinearity. It is
shown that in contrast to classical and semiclassical results \cite%
{chatterjee2015,chatterjee2016}, both the local and nonlocal terms of the
NLS equation get modified due to the multi-plasmon processes if the
dimensionless quantum parameter $H=\hbar \omega _{p}/mv_{F}^{2}$ or the
dimensionless wave number \thinspace $\hbar k/mv_{F}$ is not too small. In
the regime of short wavelengths such that multi-plasmon processes are
allowed, but still \thinspace $k<k_{cr}$ such that one-plasmon resonances
are forbidden, it is found that the three-plasmon processes play the
dominant role for wave damping due to wave-particle interaction. Moreover,
we note that the multi-plasmon process can affect the modulation of the wave
envelope in decreasing the frequency shift and increasing the energy transfer rate, as described by the contributions  of the nonlinear coefficients $Q$ and $R$ of Eq. \eqref{nls}. 
\par 
 To discuss about whether the frequency up-shift or down-shift occurs in the modulation of wave envelopes, we note that in the process of modulation of a plane wave solution [Eq.\eqref{sol-nls}] of the NLS equation [Eq. \eqref{nls}] by
plane wave perturbations   [Eqs. \eqref{rho-perturbation} and \eqref{sigma-perturbation}], the solution \eqref{sol-nls}, in fact, describes a three-wave interaction (see for details, e.g.,  Ref. \onlinecite{chatterjee2015}) of
the unperturbed carrier wave $(\omega_0,k_0)$ and two side bands with wave
numbers $k_0\pm\epsilon |K|$ and frequencies $\omega_0\pm\lambda|K|\pm%
\epsilon^2\Omega$, where $\epsilon$ is some scaling parameter and $\lambda$
is the group velocity. Here, we take the upper (lower) sign for $K>0~(K<0)$.
Thus, for $K>0$, the frequency of the carrier pump wave $\omega_0$ is
up-shifted or down-shifted according to when   $\Omega_r$ is positive or negative. In the present theory, we find that the expression for the frequency shift $\Omega_r$ changes significantly
due to the presence of the nonlocal nonlinearity  ($\propto R$ associated with the group
velocity and three-plasmon resonances) in the NLS equation. From  Eq. \eqref{Omega-real}   it is clear that $\Omega_r$ can never be
negative as it explicitly depends on $K$ but not on $P$ (the group velocity
dispersion). However, $\Omega_r$ may be zero if the carrier-wave frequency
can turn over with the group velocity dispersion going to zero and then to
negative values (see, e.g., Ref. \onlinecite{chatterjee2015}). This is not the case in our present work. On the other hand, in absence of the nonlocal nonlinearity, the expression for $\Omega$ [i.e., Eq.  \eqref{sol-nls}
with $R=0$] explicitly depends on $P$, in which case, the
frequency shift (in the case of stable wave oscillation for a certain $K>K_c$, otherwise the wave is unstable for $K<K_c$ with   $K_c$ denoting some critical
wave number of modulation) can be positive or negative depending on whether $P>0$ or $P<0$. Thus, in our present analysis, the nonlinear effects (especially the nonlocal nonlinearity) lead  $\Omega_r$ to be  positive for $K>0$, resulting into a frequency up-shift of the pump wave.  
\par
One of the earliest investigations of P. A. Sturrock \cite{sturrock1957} on nonlinear Langmuir waves may be discussed and compared with the present work. However, The work of P. A. Sturrock is mainly concerned with the coherent
and incoherent interactions of Langmuir waves in electron fluid plasmas,
especially interaction of three waves. It was shown that the coherent
interaction is responsible only for a frequency shift associated with each
wave number of the dispersion relation in which no exchange of energy
between wave numbers takes place. However, the incoherent interaction is
responsible for spectral decay in which the redistribution energy takes
place in wave number space. Here, the damping mechanism is due to the
particle-particle collisions quite distinctive from the Landau damping
mechanism (wave-particle interaction) in collisionless plasmas as in our
present theory. From our quantum kinetic theory, one can not exactly recover
the classical results, because even though the quantum effects associated
with the terms $\propto \hbar k/m$, which appear due to the Wigner equation
rather than the Vlasov equation can be negligible, however, the quantum
contribution due to the background distribution (Fermi-Dirac at zero
temperature) of electrons rather than the Maxwellian can not be ignored.
What we can say is that in the semiclassical limit, the frequency is
up-shifted and remains high, however, the rate of transfer of wave energy
from high-frequency side bands to lower ones is greatly reduced.  
\par
Furthermore, one important question may be raised in this context: Does the  group
velocity  of a plasma wave give rise to a  phase velocity  of a nonlinearly driven wave? The answer is no, i.e., not the case in our present theory. This could be an important mechanism, in
principle, e.g., for wake field generation for short pulses, similar to the well-known generation mechanism in the laser wake field scheme. However, we do not focus on the regime of very short Langmuir pulses and thereby such effects are forbidden. 
\par
The present approach may be
generalized to cover the case of a finite temperature plasma, which is a
project for future work.
\acknowledgments{This work was supported by UGC-SAP (DRS, Phase III) with 
Sanction  order No.  F.510/3/DRS-III/2015(SAPI),  and UGC-MRP with F. No.
43-539/2014 (SR) and FD Diary No. 3668.} 
\appendix
\section{Expressions for $A,~A_1,~B$ and $C$ in $\gamma$}\label{appendix-a} 
\begin{widetext} 
\begin{equation}
A=-\frac{16\pi e^3}{ A_0m^2k^3}{\int_C \frac{(v_p-v)\left[ (v_p-v)^2+%
\frac{v_q^2}{2}\right] }{\left\lbrace (v_p-v)^2-v_q^2\right\rbrace ^2 \left\lbrace \left(v_p-v+v_q%
\right)^2-v_q^2\right\rbrace \left\lbrace \left(v_p-v-%
v_q\right)^2-v_q^2\right\rbrace }%
F^{(0)}(v)dv},  \label{A}
\end{equation}
where
\begin{equation}
A_0={1-\frac{\pi e^2}{k^2m}\int_C \frac{F^{(0)}(v)}{(v_p-v)^2-\left(2v_q \right)^2 }}dv.
\end{equation}
Also, 
\begin{eqnarray}
&&A_1=-{12\pi e^3}\frac{\hbar }{k^2 m^2}\left[ \int_C \frac{1}{\left\lbrace
\left(v_p-v+v_q\right)^2-4v_q^2%
\right\rbrace \left\lbrace \left(v_p-v-v_q\right)^2-%
4v_q^2\right\rbrace dv }\right.  \notag \\
&&\left.+ \frac{3}{2}\int_C \frac{\left[ (v_p-v)^2+v_q^2\right] }{\left\lbrace (v_p-v)^2-4v_q^2\right\rbrace
\left\lbrace \left(v_p-v+2v_q\right)^2-v_q^2\right\rbrace \left\lbrace \left(v_p-v-2v_q\right)^2-%
v_q^2\right\rbrace }\right] F^{(0)}(v)dv,  \notag
\label{A1}
\end{eqnarray}
\begin{eqnarray}
&& B=\frac{4\pi e^4}{k^4 m}\int_C \left[ \frac{1}{\left\lbrace
v_p-v+2v_q\right\rbrace \left\lbrace v_p-v+v_q\right\rbrace \left\lbrace \left(v_p-v+2v_q%
\right)^2-v_q^2\right\rbrace} \right.  \notag \\
&&\left. +\frac{1}{\left\lbrace v_p-v-2v_q\right\rbrace
\left\lbrace v_p-v-v_q\right\rbrace \left\lbrace
\left(v_p-v-2v_q\right)^2-v_q^2%
\right\rbrace}\right.  \notag \\
&& \left. - \frac{2}{\left\lbrace (v_p-v)^2-v_q^2%
\right\rbrace^2 }\right]F^{(0)}(v)dv,  \notag  \label{B}
\end{eqnarray}
\begin{eqnarray}
&& C(k,\omega; \lambda)=-\frac{4\pi e^4}{m \hbar^2 k^2}
\int_C \frac{1}{(v_p-v)^2-v_q^2} \frac{I(v)}{%
v-\lambda}dv  \notag \\
&&=-\frac{4\pi e^4}{m \hbar^2 k^4} \int_C \left[ \frac{v-\lambda-v_q}{\left\lbrace \left(v_p-v+2v_q\right)^2-v_q^2\right\rbrace \left(v_p-v+v_q\right)^2 \left( v-\lambda -2v_q\right)} \right.  \notag \\
&&\left. - \frac{v-\lambda+ v_q}{\left\lbrace \left(v_p-v-2v_q\right)^2-v_q^2\right\rbrace
\left(v_p-v-v_q \right)^2 \left( v-\lambda + 2v_q\right)} \right.  \notag \\
&&\left. -2v_q \frac{ \left\lbrace \left( v_p-v\right)^2+v_q^2 \right\rbrace }{(v-\lambda)\left\lbrace \left(v_p-v\right)^2-v_q^2 \right\rbrace ^3 }-4v_q\frac{v_p-v}{\left\lbrace \left( v_p-v\right)^2-v_q^2 \right\rbrace ^3}\right] F^{(0)}(v)dv,  \notag
\label{C-k-omega}
\end{eqnarray}
where
\begin{equation}
I(v)= \frac{1}{k^2}\left[ \left(v-\lambda+v_q \right) \frac{f^{(0)}\left( v+2v_q\right) -f^{(0)}(v)}{{\left\lbrace v_p-\left(v+v_q \right) \right\rbrace }^2} +\left(v-\lambda-v_q \right) 
\frac{f^{(0)}(v)-f^{(0)}\left( v-2v_q\right) }{{%
\left\lbrace v_p-\left(v-v_q \right) \right\rbrace }^2}\right].\label{I-v}
\end{equation}
\section{Reduced expressions for $\alpha,~\beta,~\gamma $ and $D$ with the Fermi distribution at zero temperature }\label{appendix-b} 
\begin{equation}
\alpha=-\frac{8m \omega_p^2}{3\hbar k^2 v_F^3} \sum_{j=\pm1} \left(v_p+jv_q
\right) \log \left\vert\frac{v_p+jv_q-v_F}{v_p+jv_q+v_F}\right\vert ,
\label{alpha-eq}
\end{equation}
\begin{equation}
\begin{split}
&\beta=1-\frac{3 m \omega_p^2}{2 \hbar k^3 v_F^2} \sum_{j=\pm1}\left[
\left\lbrace 2\left(v_p+jv_q \right)+\left( v_p-\lambda\right)\right\rbrace
\log \left\vert\frac{v_p+jv_q-v_F}{v_p+jv_q+v_F}\right\vert \right. \\
&\left.-\left\lbrace v_F^2-\left(v_p+jv_q \right)^2-2\left(v_p+jv_q
\right)\left(v_p-\lambda\right) \right\rbrace \frac{v_F}{ v_F^2-%
\left(v_P+jv_q \right)^2} +\left(v_p-\lambda \right)\frac{v_F \left(v_p+jv_q
\right) }{ v_F^2-\left(v_p+jv_q \right)^2}\right],  \label{beta-coeff}
\end{split}%
\end{equation}
\begin{equation}
\gamma=\left(\frac{1}{4}\frac{A A_1}{\hbar}-\frac{1}{2\hbar^2}B+C\right)k^2,
\label{gamma}
\end{equation}
where
\begin{equation}
\begin{split}
&A= - \frac{4em^2 \omega_p^2}{A_0\hbar^3 k^6 v_F^3}\sum_{j=\pm1}\left[
kv_Fv_q +\frac{\omega_p}{6v_q} \left\lbrace \left(v_F^2-\left(v_p+jv_q
\right)^2 \right)\left(4-\frac{jk v_q}{2 \omega_p} \right)\right.\right.
\\
&\left.\left. -6v_q\left(v_p+jv_q \right) \left(1-\frac{j k v_q}{%
2 \omega_p} \right)\right\rbrace \log \left\vert\frac{v_p+jv_q-v_F}{%
v_p+jv_q+v_F}\right\vert\right. \\
&\left.+\frac{\omega_p}{3 v_q} \left\lbrace v_F^2-\left(v_p+j2v_q
\right)^2 \right\rbrace \left(1-\frac{jk v_q}{4\omega_p} \right)\log
\left\vert\frac{v_p+j2v_q-v_F}{v_p+j2v_q+v_F}\right\vert\right. \\
&\left. +i \pi v_q \omega_p \left\lbrace v_F^2-\left(v_p-2v_q
\right)^2 \right\rbrace \left(1+\frac{k v_q}{4\omega_p} \right)\right],
\label{A-reduced}
\end{split}%
\end{equation}
with
\begin{equation}
\begin{split}
& A_0=1-\frac{3\omega_p^2}{16 v_F^2 k^2}\left[2-\sum_{j=\pm1}\frac{jv_q}{%
4 v_F}\left\lbrace v_F^2-\left(v_p+jv_q\right)^2 \right\rbrace \log
\left\vert\frac{v_p+j2v_q-v_F}{v_p+j2v_q+v_F}\right\vert \right] \\
&-i\frac{3\pi  \omega_p^2}{64 v_F^3 v_q k^2}\left\lbrace
v_F^2-\left(v_p-2v_q \right)^2 \right\rbrace.  \label{A-0-reduced}
\end{split}%
\end{equation}
Also, 
\begin{equation}
\begin{split}
&A_1=-\frac{e}{m}\frac{9\omega_p^2 m^3}{4\hbar^2 v_F^3k^5} \sum_{j=\pm1}%
\left[\frac{j}{4}\left\lbrace v_F^2-\left(v_p+jv_q\right)^2 \right\rbrace
\log \left\vert\frac{v_p+jv_q-v_F}{v_p+jv_q+v_F}\right\vert\right. \\
&\left.- \frac{2j}{3}\left\lbrace v_F^2-\left(v_p+j3v_q \right)^2
\right\rbrace \log \left\vert\frac{v_p+j3v_q-v_F}{v_p+j3v_q+v_F}\right\vert
+j\left\lbrace v_F^2-\left(v_p+j2v_q \right)^2 \right\rbrace \log \left\vert%
\frac{v_p+j2v_q-v_F}{v_p+j2v_q+v_F}\right\vert\right] \\
& -i\frac{9\pi \omega_p^2}{16v_F^3k^3}\frac{e }{v_q^2}\left[ \frac{2}{3}%
\left\lbrace v_F^2-\left(v_p-3v_q \right)^2 \right\rbrace-\left\lbrace
v_F^2-\left(v_p-2v_q \right)^2 \right\rbrace\right],
\end{split}%
\end{equation}
\begin{equation}
\begin{split}
&B=-\frac{e^2}{k^7}\frac{3\omega_p^2m^3}{2v_F^3\hbar^3}\sum_{j=\pm1}\left[ 
16 v_F v_q+4\left\lbrace v_F^2-\left(v_p+j2v_q\right)^2
\right\rbrace \log \left\vert\frac{v_p+j2v_q-v_F}{v_p+j2v_q+v_F}%
\right\vert\right. \\
&\left.+\left\lbrace v_F^2-\left(v_p+j3v_q\right)^2 \right\rbrace \log
\left\vert\frac{v_p+j3v_q-v_F}{v_p+j3v_q+v_F}\right\vert -\left\lbrace
v_F^2-\left(v_p+jv_q \right)^2 -8j v_q\left( v_p+jv_q\right)
\right\rbrace \log \left\vert\frac{v_p+jv_q-v_F}{v_p+jv_q+v_F}\right\vert %
\right] \\
& -i \frac{3 \pi e^2 \omega_p^2}{4 k^4 v_F^3}\left[-\frac{1}{ v_q^3}%
\left\lbrace v_F^2-\left(v_p-2v_q\right)^2 \right\rbrace + \frac{1}{%
4 v_q^3}\left\lbrace v_F^2-\left(v_p-3v_q \right)^2 \right\rbrace\right],
\label{B-reduced}
\end{split}%
\end{equation}
\begin{equation}
\begin{split}
& C=-\frac{3}{4} \frac{e^2}{\hbar^2 k^4}\frac{\omega_p^2}{v_F^3}%
\sum_{j=\pm1} \left[ -\frac{1}{8 v_q^3} \frac{v_p+j2v_q-\lambda}{%
v_p+jv_q-\lambda} \left\lbrace v_F^2-\left(v_p+j3v_q \right)^2 \right\rbrace
\log \left\vert\frac{v_p+j3v_q-v_F}{v_p+j3v_q+v_F}\right\vert +jM_j \log
\left\vert\frac{v_p+jv_q-v_F}{v_p+jv_q+v_F}\right\vert\right. \\
&\left. +jN_j \frac{2v_F}{v_F^2-\left(v_p+jv_q \right)^2 } -\left( \frac{1}{%
2 v_q} \frac{v_p-\lambda}{v_p-jv_q-\lambda} -\frac{1}{2} \frac{1}{%
v_p+jv_q-\lambda}-\frac{1}{2 v_q}\right) \frac{2v_F\left(v_p+jv_q \right) 
}{v_F^2-\left(v_p+jv_q\right)^2 }\right. \\
& \left. -v_q \frac{v_F^2-\left(\lambda +jv_q\right)^2 }{\left(v_p-jv_q-%
\lambda \right)^3 \left( v_p+jv_q-\lambda\right) }\log\left\vert \frac{%
\lambda +jv_q-v_F}{\lambda+jv_q+v_F}\right\vert +\frac{2v_q}{k^2} \frac{%
\left\lbrace \left( \omega-k \lambda\right)^2 +\frac{\hbar^2 k^4}{4m^2}
\right\rbrace\left( v_F^2-\lambda^2\right) }{\left(v_p+v_q-\lambda \right)^3
\left( v_p-v_q-\lambda\right)^3} log \frac{\lambda-v_F}{\lambda+v_F}\right],
\end{split}
\label{C-reduced}
\end{equation}
with
\begin{equation}
\begin{split}
&M_{1,-1}= \frac{1}{4v_q^2 \left(v_p-\lambda\mp 2v_q \right)^2 }\left[%
\left(v_p-\lambda\mp 3v_q \right)\left\lbrace v_F^2-3\left( v_p\pm
v_q\right)^2+2\left(\lambda\pm v_q \right)\left( v_p\pm
v_q\right)\right\rbrace \right. \\
& \left. \mp4v_q\left(v_p-\lambda\mp v_q \right) \left(3v_p-\lambda\pm 2v_q
\right) -2 \left(v_p\pm v_q \right)\left( v_p-\lambda\right)
\left(v_p-\lambda\mp 3v_q \right)\right. \\
&\left.+\left\lbrace 2\left(v_p-\lambda\mp v_q \right)+\left(v_p-\lambda\mp
3v_q \right)\pm 2\left(v_q-\lambda \right)\left(v_q-\lambda\mp3v_q
\right)^2\left(\frac{1}{2v_q}\mp\frac{1}{v_p- \lambda\mp 2v_q} \right)
\right\rbrace \left\lbrace v_F^2-\left(v_p\pm v_q \right)^2 \right\rbrace %
\right] \\
& +\frac{1}{4v_q^2 \left(v_p- \lambda\pm v_q \right)^2} \left[
5\left(v_p-\lambda\pm v_q \right) \left\lbrace v_F^2-\left(v_p\pm v_q
\right)^2 \right\rbrace \pm 3v_q \left\lbrace v_F^2-\left(v_p\pm
v_q\right)^2 \right\rbrace \right. \\
&\left. \mp 4v_q^2\left( v_p \pm v_q\right) -\left(3v_p-3\lambda\pm 5v_q
\right) \left\lbrace v_F^2\mp 2v_q\left(v_p\pm v_q \right)- \left(v_p\pm v_q
\right)^2 \right\rbrace \right. \\
& \left.+2v_q^2\frac{v_F^2-\left(v_p\pm v_q\right)^2 }{v_p-\lambda\pm v_q}%
\right] \pm \frac{v_p}{v_q^2},
\end{split}%
\end{equation}
\begin{equation}
\begin{split}
& N_{1,-1}=\pm \frac{1}{4v_q^2 \left(v_p-\lambda\mp v_q \right)^2}\left[%
2v_q\left(v_p-\lambda\mp v_q \right)\left\lbrace v_F^2-3\left(v_p\pm
v_q\right)^2 -2\left(v_p-\lambda\right) \left(v_p\pm v_q \right)
\right\rbrace \right. \\
&\left.+\left\lbrace v_F^2-\left( v_p\pm v_q \right)^2 \right\rbrace
\left(v_p-\lambda\right)\left(v_p-\lambda\mp 3v_q \right) \right] \\
&\mp\frac{1}{4v_q{\left(v_p-\lambda\mp v_q \right)^2}}\left[
\left(v_p-\lambda\mp 3v_q \right)\left\lbrace v_F^2-\left(v_p\pm v_q
\right)^2 \right\rbrace \pm 4v_q\left(v_p\pm v_q \right)\left(v_p-\lambda\pm
v_q \right)\right] \\
&\pm \frac{1}{4v_q^2}\left[v_p\pm v_q- \left\lbrace v_F^2-\left(v_p\pm v_q
\right)^2 \right\rbrace \right],
\end{split}%
\end{equation}
\end{widetext} \begin{widetext} 
\begin{equation}
D=\frac{3 e^2 \pi \omega_p^2}{4 k\hbar m v_F^3}\left[(v_F^2-\lambda^2) \frac{(v_p-\lambda)^2+v_q^2}{{\left\lbrace(v_p-%
\lambda)^2-v_q^2 \right\rbrace^3 }} +\frac{1}{8 v_q^4}\left\lbrace
v_F^2 -\left(v_p-3v_q \right)^2 \right\rbrace \frac{ v_p-\lambda-2v_q}{%
v_p-\lambda-v_q}\right].  \label{D-reduced}
\end{equation}
\end{widetext} This expression \eqref{D-reduced} of $D$ is obtained by using
the following relations. 
\begin{eqnarray}
&& \lim_{\nu _g\rightarrow 0} \frac{1}{\Omega-Kv+i\nu_g}=\frac{1}{\Omega-Kv}%
-i\pi\frac{1}{|K|}\delta \left(v-\frac{\Omega}{K} \right),  \notag \\
&& \lim_{\nu _3\rightarrow 0} \frac{1}{\omega-kv-3kv_q+i\nu_3%
}= \frac{1}{\omega-kv-3kv_q}  \notag \\
&&-i\pi\frac{1}{|K|} \delta\left(v-v_p+3v_q
\right),
\end{eqnarray}
and we have made use of $\Omega/K\rightarrow\lambda$. The infinitesimal
quantities $|\nu_g|$ and $|\nu_3|$ are taken to anticipate the Landau
damping terms associated with the group velocity and three-plasmon
resonances. 
\par
Thus, the reduced expressions of $P,~Q$ and $R$ can be obtained from the relations $P=\beta/\alpha$, $Q=\gamma/\alpha$ and $R=D/\alpha$.


\begin{thebibliography}{99}

\bibitem{Nicholsson} D. R. Nicholson, \textit{Introduction to Plasma Theory, 
}(John Wiley \& Sons, New York, 1983).
\bibitem{Manfredi-1997} G. Manfredi, Phys. Rev. Lett. \textbf{79}, 2815
(1997).
\bibitem{Danielson2004} J. R. Danielson, F. Anderegg, and C. F. Driscoll,
Phys. Rev. Lett. \textbf{92}, 245003 (2004).
\bibitem{rightley2016} S. Rightley and D. Uzdensky, Phys. Plasmas \textbf{23}%
, 030702 (2016).
\bibitem{mendonca2016} J. T. Mendonca and A. Serbeto, Phys. Scr. 
\textbf{91}, 095601 (2016).
\bibitem{daligault2014} J. Daligault, Phys. Plasmas \textbf{21}, 040701
(2014).
\bibitem{misra2011} A. P. Misra and S. Banerjee, Phys. Rev. E \textbf{83}, 037401 (2011).
\bibitem{ikezi1971} H. Ikezi and Y. Kiwamoto, Phys. Rev. Lett. \textbf{27}, 718 (1971).
\bibitem{chatterjee2015} D. Chatterjee and A. P. Misra, Phys. Rev. E \textbf{%
92}, 063110 (2015).
\bibitem{chatterjee2016} D. Chatterjee and A. P. Misra, Phys. Plasmas 
\textbf{23}, 102114 (2016).
\bibitem{brodin2016} G. Brodin, R. Ekman, and J. Zamanian, Plasma Phys. Control. Fusion (2017), doi: 10.1088/1361-6587/aa979d.
\bibitem{brodin2017} G. Brodin,   J.  Zamanian, and J. T. Mendonca, Phys. Scr. \textbf{90}, 068020 (2015). 
\bibitem{Ekman 2015} R. Ekman, J. Zamanian and G. Brodin, Phys. Rev. E 
\textbf{92}, 013104 (2015).
\bibitem{eliasson2010} B. Eliasson and P. K. Shukla, J. Plasma Phys. \textbf{%
76}, 7 (2010). 
\bibitem{krivitskii1991} V. S. Krivitskii and S. V. Vladimirov, SOv. Phys. J. Exp. Th. Phys. (JETP) \textbf{73}, 821 (1991).
\bibitem{ichikawa1974} Y. H. Ichikawa, Suppl. Prog. Theor. Phys. \textbf{55}, 212 (1974).
\bibitem{ferrel1957} R. A. Ferrel, Phys. Rev. \textbf{107}, 450 (1957).
\bibitem{sturrock1957} P. A. Sturrock, Proc. R. Soc. Lond. A \textbf{242}, 277 (1957).

%
\end{thebibliography}
\end{document}